\documentclass[journal=jctcce , layout=traditional,manuscript=article]{achemso}
\SectionNumbersOn
\setkeys{acs}{articletitle = true}

\usepackage{chemformula} 
\usepackage[T1]{fontenc} 
\usepackage{amsmath}%
\usepackage{MnSymbol}%
\usepackage{wasysym}%
\usepackage{ulem}%
\usepackage{tabularx}
\usepackage{caption}
\usepackage{achemso}
\setkeys{acs}{maxauthors = 0}
\usepackage{CJK}

\usepackage{siunitx}

\usepackage{bm}%
\usepackage[colorlinks=true,linkcolor=blue]{hyperref}%
\usepackage{graphicx}
\usepackage[utf8]{inputenc}
\usepackage{epstopdf}
\usepackage{afterpage}
\usepackage{setspace}
\usepackage{natbib}
\usepackage[none]{hyphenat}
\usepackage{blindtext}
\usepackage[symbol]{footmisc}
\usepackage{float}
\usepackage{multirow}

\usepackage[caption=false,justification=centerlast]{subfig}

\DeclareUnicodeCharacter{0301}{*}
\DeclareUnicodeCharacter{2A7D}{$\le$}

\newcommand{\rvline}{\hspace*{-\arraycolsep}\vline\hspace*{-\arraycolsep}}
\newcommand{\erf}[1]{\mathrm{erf}\left(#1\right)}

\newcommand{\M}[1]{\mathbf{#1}}

\expandafter\ifx\csname package@font\endcsname\relax\else
 \expandafter\expandafter
 \expandafter\usepackage
 \expandafter\expandafter
 \expandafter{\csname package@font\endcsname}%
\fi
\title[]{Accurate Fourth-Generation Machine Learning Potentials by Electrostatic Embedding}

\author{Tsz Wai Ko}
\email{tko@chemie.uni-goettingen.de}
\affiliation{Universit\"{a}t G\"{o}ttingen, Institut f\"{u}r Physikalische Chemie, Theoretische Chemie, Tammannstra\ss{}e 6, 37077 G\"{o}ttingen, Germany}

\author{Jonas A. Finkler}

\affiliation{Department of Physics, Universit\"{a}t Basel, Klingelbergstrasse 82, 4056 Basel, Switzerland}
\author{Stefan Goedecker}
\affiliation{Department of Physics, Universit\"{a}t Basel, Klingelbergstrasse 82, 4056 Basel, Switzerland}

\author{J\"{o}rg Behler}
\email{joerg.behler@ruhr-uni-bochum.de}
\affiliation{Universit\"{a}t G\"{o}ttingen, Institut f\"{u}r Physikalische Chemie, Theoretische Chemie, Tammannstra\ss{}e 6, 37077 G\"{o}ttingen, Germany}
\affiliation{Present Address: Lehrstuhl f\"ur Theoretische Chemie II, Ruhr-Universit\"at Bochum, 44780 Bochum, Germany, and Research Center Chemical Sciences and Sustainability, Research Alliance Ruhr, 44780 Bochum, Germany}%
\date{\today}

\begin{document}

\begin{abstract}
In recent years, significant progress has been made in the development of machine learning potentials (MLPs) for atomistic simulations with applications in many fields from chemistry to materials science. While most current MLPs are based on environment-dependent atomic energies, the limitations of this locality approximation can be overcome, e.g., in fourth-generation MLPs, which incorporate long-range electrostatic interactions based on an equilibrated global charge distribution. Apart from the considered interactions, 
the quality of MLPs crucially depends on the information available about the system, i.e., the descriptors. 
In this work we show that including --- in addition to structural information --- the electrostatic potential arising from the charge distribution in the atomic environments significantly improves the quality and transferability of the potentials. Moreover, the extended descriptor allows to overcome current limitations of two- and three-body based feature vectors regarding artificially degenerate atomic environments.
The capabilities of such an electrostatically embedded fourth-generation high-dimensional neural network potential (ee4G-HDNNP), which is further augmented by pairwise interactions, are demonstrated for NaCl as a benchmark system. Employing a data set containing only neutral and negatively charged NaCl clusters, even small energy differences between different cluster geometries can be resolved, and the potential shows an impressive transferability to positively charged clusters as well as the melt.
\end{abstract}

\maketitle 

\section{Introduction}

Machine learning potentials (MLPs) have become an increasingly important tool for performing atomistic simulations since they offer a unique combination of close-to ab initio accuracy and force field-like computational efficiency~\cite{P4885,P5673,P6121,P5793,P6131,P6112}. 
Most MLPs rely on the assumption that a large part of the atomic interactions is determined by the local structure. Based on this ansatz the total energy of the system can be constructed as a sum of atomic energy contributions~\cite{P1174}, which in turn are functions of the local atomic environments within a spatial cutoff radius $R_{\mathrm{c}}$. Local MLPs such as high-dimensional neural network potentials (HDNNPs)~\cite{P1174,P6018}, Gaussian approximation potentials (GAPs)~\cite{P2630}, spectral neighbor analysis potentials (SNAPs)~\cite{P4644}, atomic cluster expansion (ACE)~\cite{P5794}, moment tensor potentials (MTPs)~\cite{P4862} and many others have been successfully used in numerous applications in physics, chemistry and materials science. 
\par
Often, the local chemical environments are represented by structural descriptors with predefined functional forms ensuring the mandatory translational, rotational and permutational invariances of the atomic energies. Alternatively, the description of the system can be included in the learning process making use of adjustable atomic feature vectors, which are repeatedly updated by propagating structural and other information in message passing neural networks (MPNNs)~\cite{P5368}. Various MLPs of the latter form such as DTNN~\cite{P4937}, SchNet~\cite{P5366}, HIPNN~\cite{P5888}, and AIMNET~\cite{P5817} have been proposed in recent years. 

\par
A main limitation of local MLPs is the lack of long-range interactions beyond the cutoff radius, or beyond the maximum information exchange distance in message passing networks, most importantly electrostatics, which prevents the study of many interesting phenomena from biological processes~\cite{sagui1999molecular,ren2012biomolecular} to electrochemical interfaces~\cite{sundararaman2022improving,magnussen2019toward,onofrio2019exploring}. A straightforward solution is to include long-range electrostatic contribution explicitly employing, e.g., fixed atomic charges~\cite{P2630,P5614} or flexible environment-dependent charges expressed by machine learning~\cite{P2391,P2962,P3132,P5577,P5313,P5885}. Other methods such as the long distance equivariant (LODE)~\cite{P5629} representation can also describe the long-range electrostatic interaction implicitly based on a Coulomb-type atomic density potential. 
Still, methods based on environment-dependent, i.e., local, atomic charges cannot describe long-range charge transfer~\cite{P6156,P5977} and multiple charge states of a system, as this would require knowledge about the structure and global charge of the entire system.
\par 
Recently, a few MLPs of the fourth generation~\cite{P5977} have been developed to overcome these limitations. The first MLP being able to describe non-local phenomena has been the charge equilibration via neural network technique (CENT)~\cite{P4419}, which is based on a charge equilibration scheme~\cite{P1448} relying on atomic electronegativities, hardnesses and Gaussian charge densities to enable the global redistribution of charges over the entire system by minimizing a charge-dependent total energy including electrostatics. In the framework of CENT, the electronegativity is assumed to be a local property and is expressed by atomic neural networks (NNs). Due to its charge-based energy function CENT reaches the highest accuracy for systems with primarily ionic bonding~\cite{P4990,P5864}. 
\par 
Another approach is the Becke population neural network (BpopNN)~\cite{P5859}, in which the charges are adapted in a self-consistency loop, and the atomic energies depend on both the structure and the atomic populations. The accuracy of the method, which in addition to atomic energies also includes Coulomb interactions, has been demonstrated for lithium hydride clusters.
\par 
More recently, fourth-generation high-dimensional neural network potentials (4G-HDNNPs)~\cite{P5932,P5977}, which combine the advantages of CENT and HDNNPs, have been proposed. Like in CENT, the global charge distribution is determined in a charge equilibration step, which is relying on local environment-dependent electronegativities expressed by atomic NNs. In contrast to CENT, the resulting partial charges are trained to reference charges. They are not only used to calculate the electrostatic energy but also --- in combination with atom-centered symmetry functions~\cite{behler2011atom} --- as descriptors, which are supplied to atomic NNs yielding the short-range atomic energies. Thus, the atomic energies consider possible non-local charge transfer and its implications for local bonding. The reliability of 4G-HDNNPs, which like CENT and BpopNN are also applicable to multiple charge states of a system, has been demonstrated for different types of model systems from molecules to periodic surfaces~\cite{P5932}. 
\par 
Next to these fourth-generation MLPs employing predefined descriptors, also MPNNs have been used to construct MLPs for systems including dependencies beyond the local atomic environments. For instance, AIMNET-NSE~\cite{zubatyuk2021teaching} allows to distinguish different charge states by minimizing the error between predicted and reference charges during the iterative refinement of the atomic feature vectors. Atomic weight factors, which can be conceptually interpreted as atomic Fukui functions, are used to redistribute the charges over larger distances depending on the number of message passing steps.
Another model is QRNN~\cite{jacobson2022transferable}, which is essentially a simplified charge equilibration scheme for determining the global charge distribution. It employs local effective electronegativities implicitly including the off-diagonal terms in the Coulomb interaction matrix.
Therefore the charges can be calculated in an approximate way without solving a system of linear equations of the conventional charge equilibration. 
Further, SpookyNet~\cite{P6056}, which takes the nuclear charges, the number of electrons and the spin states as atomic feature representations, allows to capture long-range dependencies in the electronic structure.
The electronic information, which is distributed across atoms with atomic weights, are iteratively exchanged through message passing. 
Another MPNN-based method for determining charge distributions is the electron passing neural network~\cite{P5933}.
\par
In summary, modern MLPs are characterized by including more and more physical information, starting from completely local potentials making use of environment-dependent atomic energies only, via MLPs including long-range electrostatics employing element-specific or environment-dependent charges, to MLPs capturing non-local phenomena like long-range charge transfer.
\par
In particular for methods relying on predefined descriptors, the information included in the descriptors is important for the construction of high-quality PESs. While to date most descriptors provide structural information, only recent methods, like BpopNN, 4G-HDNNP, and QRNN include further information like atomic charges in the atomic energy terms. In a similar way, also numerous MPNNs make use of charges, but the iterative determination of atomic feature vectors, although more convenient, is often less transparent and less efficient than the use of predefined descriptors.
\par
In spite of these advances, the accuracy of current MLPs is still limited by their ability to distinguish different bonding situations, which requires sufficient information about the system. Recent studies~\cite{pozdnyakov2020incompleteness, parsaeifard2022manifolds} have shown that commonly used atomic environment descriptors such as atom centered symmetry functions (ACSFs)~\cite{behler2011atom} and the smooth overlap of atomic positions~\cite{bartok2013representing} provide an incomplete description of the atomic environment due to the lack of higher order terms. As a results, in rare cases different atomic environments can map to the exact same descriptor values, which lead to the atomic NN predicting the same atomic energy.  Although such a degeneracy does not generally affect the accuracy of MLPs for condensed systems it could yield an inaccurate description for small molecules.
4G-HDNNPs may be able to break the artificial degeneracy by taking the charge transfer arising from the change in the atomic environments into account. 
Furthermore, 4G-HDNNPs allow to study systems with different global charges but identical geometries. This is achieved by including the atomic partial charge of the respective atom as additional input neuron in the atomic NNs yielding the short-range energies, which thus depend on the global structure.
\par
\begin{figure}
    \centering
    \includegraphics[width=0.5\textwidth]{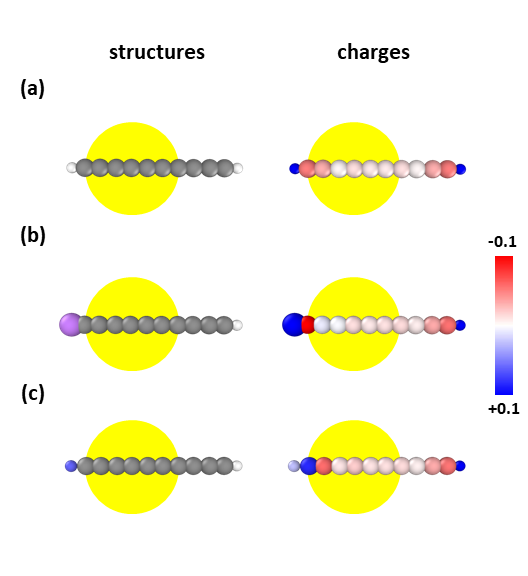}
    \caption{\textbf{A molecular system illustrating the importance of capturing electronic information in the atomic environments.} Panel (a) shows the structure of C$_{10}$H$_{2}$ optimized using density-functional theory (DFT) and the corresponding DFT Hirshfeld charges~\cite{P0416} (in units of the elementary charge $e$) as a qualitative visualization of the electronic structure. As reference atom we choose the fourth carbon atom from the left and highlight its local chemical environment using a typical cutoff radius by a yellow sphere. Replacing the hydrogen atom on the left by a lithium atom in (b) (the C-Li bond has been relaxed) substantially changes the electronic structure of the left half of the molecule, while the local geometric environment of the reference carbon atom is still identical to panel (a), and also its atomic charge is very similar while the neighboring atomic charges clearly differ. Substituting the same hydrogen atom by a fluorine atom in (c) (the C-F bond has been relaxed) has a similar effect but with the partial charges between the fluorine and the reference carbon now changing in the opposite way. In all these cases, the identical local geometry and very similar atomic charge of the reference carbon atom will yield essentially the same atomic energy in a 4G-HDNNP, which is limiting the accuracy of the potential because the modified electronic structure changes the atomic interactions. 
    On the left, the hydrogen, lithium, carbon and fluorine atoms are colored in white, purple, grey, and blue, respectively. The structure has been visualized using Ovito~\cite{P5515}.}
    \label{fig:example_4G-PINNP}
\end{figure}

However, there might still be situations in which two atoms in two different systems are embedded in a different electronic structure while both, the atomic partial charges and geometric environments, are very similar. In such a situation this information is still insufficient and the accuracy of the PES will be reduced due to contradictory training data. A small molecule illustrating such a case is shown in Fig.~\ref{fig:example_4G-PINNP}. 
While in this simple one-dimensional system the origin of contradictory training data and the resulting limited accuracy of MLPs is straightforward to analyze, realistic training sets of complex systems typically consist of thousands of structures, each containing hundreds of atoms, and in such situations inconsistent data is likely to remain undetected.
This raises the question if the explicit inclusion of additional physically interpretable descriptors would allow to further improve the quality of MLPs.
\par
Another well-known limitation of current MLPs is the very limited transferability to local atomic environments outside the training set. 
Several attempts have been made to address this problem such as the combination of empirical pairwise potentials and MLPs~\cite{dolgirev2016machine}, or introducing two-body and three-body descriptors for capturing explicit two- and three-body contributions~\cite{rowe2020accurate,rowe2018development}. 
These methods can indeed improve the transferability of MLPs by ensuring repulsive forces when atoms are very close to each other even if such local environments are not covered in the training set.
\par
In the present work, we address both of these challenges. In a first step we augment 4G-HDNNPs by explicitly including empirical two-body interaction terms taking Pauli repulsion and van-der-Waals-like interactions into account. 
This improves the reliability in case of extrapolation and thus the transferability of the potential. 
\par
In a second step we propose an electrostatically embedded fourth-generation high-dimensional neural network potential (ee4G-HDNNP), which contains a set of additional physically interpretable descriptors serving as additional inputs for the short-range atomic NNs.
These descriptors correspond to the element-resolved electrostatic potentials acting on the central atom arising from all neighboring Gaussian charges within the atomic environment up to the cutoff radius. 
These charges are identical to those obtained from the 4G-HDNNP charge equilibration step, which has been demonstrated in previous work to provide an accurate charge distribution of the entire system~\cite{P5932}.
We note that the electrostatic interaction with neighboring charges is already taken into account by the explicit Coulomb energy contribution, and here the electrostatic potential input information in the short-range part is not bound to the functional form of Coulomb's law. It thus enables a very flexible description of local covalent bonding in response to electronic structure changes. This allows to represent even challenging potential-energy surfaces. 
\par
To illustrate the performance of the ee4G-HDNNP method, we have chosen sodium chloride clusters of different size, geometry and total charge. In NaCl, which has previously been studied by MLPs~\cite{liang2021theoretical,tovey2020dft,li2021development}, electrostatic interactions play an important role calling for a MLP explicitly including these interactions. However, we will show below that the overall accuracy also strongly depends on the short-range atomic energies, which can be improved by descriptors including additional information next to atomic positions. Specifically, we will assess the impact of both extensions by comparing the accuracy of 4G-HDNNPs, 4G-HDNNPs including two-body terms (4G-HDNNP+$E_{\mathrm{2b}}$), and ee4G-HDNNPs trained to DFT data for both neutral and negatively charged sodium chloride clusters. Moreover, we include a third-generation HDNNP (3G-HDNNP), which contains electrostatic interactions based on local environment-dependent charges represented by atomic NNs~\cite{P2962} in our comparison.
This method is expected to work well for this type of ionic systems due to a clear correlation between global charge and stoichiometry.
After the analysis of several properties of these potentials for charged and neutral clusters, we perform minima hopping~\cite{goedecker2004minima} simulations driven by the ee4G-HDNNP to explore the local minima of clusters of various sizes and total charges. This represents a stringent test due to the very small energy differences between different local minima. 
Finally, we examine the transferability of the ee4G-HDNNP from clusters to periodic systems for an NaCl crystal and compare physical properties of the melt obtained from molecular dynamic simulations with DFT calculations.
%
\section{Methods} 
\subsection{Electrostatically Embedded 4G-HDNNP}

\begin{figure*}
    \centering
    \includegraphics[scale=1.0]{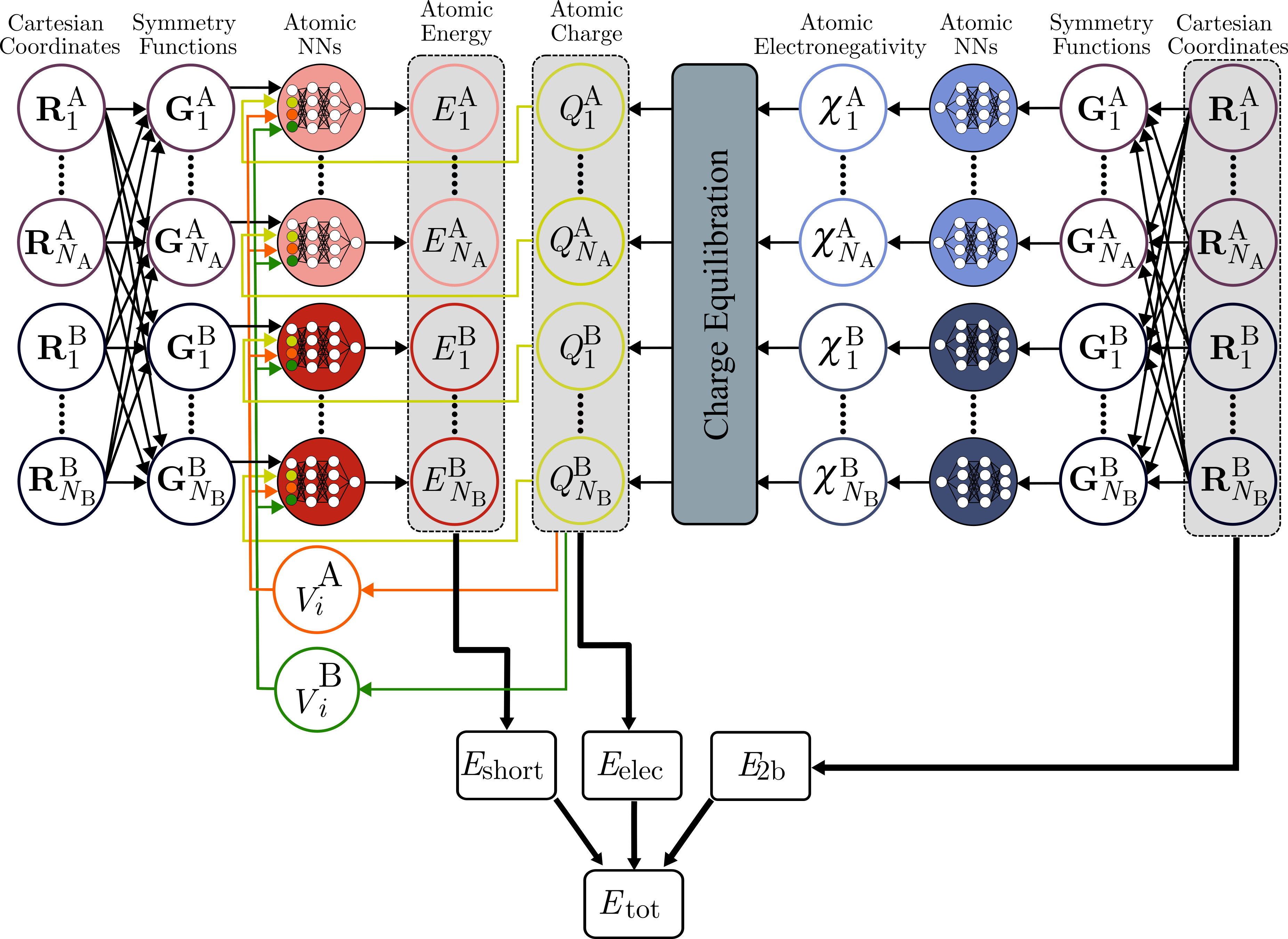}
    \caption{\textbf{Structure of an ee4G-HDNNP for a binary system.} The total energy of the system containing $N_{\mathrm{A}}$ atoms of element A and $N_{\mathrm{B}}$ atoms of element B consists of the short-range energy $E_{\mathrm{short}}$, which is a sum of atomic energy contributions $E_{i}^{\alpha}$ with $\alpha=\{A,B\}$ obtained from the atomic energy NNs, the long-range electrostatic energy $E_{\mathrm{elec}}$ calculated using Gaussian broadened charges $Q_{i}^{\alpha}$, and the two-body energies $E_{\mathrm{2b}}$. The atomic charges are globally distributed using a charge equilibration scheme relying on local environment-dependent atomic electronegativities $\chi_{i}^{\alpha}$ expressed by atomic electronegativity NNs. These charges are then used for three purposes: first to calculate the electrostatic energy, second as additional input neuron of the respective short-range atomic NN yielding the atomic energies $E_{i}^{\alpha}$, and third to compute effective element-specific electrostatic potentials $V_{i}^{\alpha}$, which are used as additional input neurons of the short-range atomic NNs. The input vectors $\mathbf{G}_{i}^{\alpha}$ of all atomic NNs describing the geometric structure of the atomic environments consist of atom-centered symmetry functions, which are obtained from the atomic Cartesian coordinates $\mathbf{R}_{i}^{\alpha}$.}
    \label{fig:extented_4G}
\end{figure*}

The starting point of our developments is the 4G-HDNNP~\cite{P5932}, which relies on atomic charges obtained in a charge equilibration step as a function of the global structure and net charge of the system. For this purpose environment-dependent atomic electronegativities are expressed by atomic NNs, which are trained to reproduce atomic reference charges. The atomic charges are not only used to compute the long-range electrostatic energy but also serve as additional input neuron for the neural network of the respective atom to compute its short-range atomic energy. In this way the atomic energies can adjust to global changes in the electronic structure arising from non-local phenomena like long-range charge transfer. The sum of all atomic energies defines the short-range energy, which represents local bonding and yields --- together with the electrostatic energy --- the potential energy of the system.
\par
In a first step, we now extend this potential for all atom pairs by two-body terms inspired by the Tosi-Fumi model~\cite{tosi1964ionic,anwar2003calculation,lee2021comparative}, which yield the total two-body energy 
\begin{equation}
    E_{\mathrm{2b}}=\sum_{i>j}^{N_{\mathrm{atoms}}}\bigg(A_{ij}^{\alpha}e^{B_{ij}^{\alpha}(\mu_{ij}^{\alpha}-R_{ij})}-\frac{C_{ij}^{\alpha}}{R_{ij}^{6}}-\frac{D_{ij}^{\alpha}}{R_{ij}^{8}}\bigg)\cdot f_{\mathrm{cut}}(R_{ij}) \label{eq:2body}
\end{equation}
covering short-range Pauli repulsion and long-ranged dispersion interactions in an empirical way. The aim is to improve the transferability of the potential to structures which are very different from the training data, and to improve the stability of the potential in case of close atomic encounters.  
$A_{ij}^{\alpha},B_{ij}^{\alpha},\mu_{ij}^{\alpha},C_{ij}^{\alpha}$ and $D_{ij}^{\alpha}$ are element-pair-specific parameters that are determined by fitting the binding energy curves of the dimers Na-Na, Na-Cl and Cl-Cl as described in the supplementary information. Further, 
\begin{equation}
         f_{\mathrm{cut}}(R_{ij}) = 
        \begin{cases}
            \mathrm{tanh}^{3}(1) & \ R_{ij} \le R_{\mathrm{c,in}} \\
            \mathrm{tanh}^{3}(1-\frac{R_{ij}-R_{\mathrm{in,c}}}{R_{\mathrm{c,2b}}-R_{\mathrm{in,c}}}) & \  R_{\mathrm{in,c}}\le R_{ij} \le R_{\mathrm{c,2b}} \\
            0 & \ R_{ij} \ge R_{\mathrm{c,2b}}
        \end{cases}
\end{equation}
is a smooth cutoff function ensuring that the two-body energies and corresponding forces decay to zero at the two-body cutoff radius $R_{\mathrm{c,2b}}$ chosen here to be 9~\si{\angstrom}, which is larger than the typical cutoff radius of 12 Bohr $\approx$ 6.35~\AA{} of the atomic environments used for the short-range atomic energies and electronegativities in the 4G-HDNNP. The inner cutoff $R_{\mathrm{c,in}}$ was set to 2.4 Bohr $\approx$ 1.27~\AA{}. A good choice of the cutoff radius for describing short-range interactions can be determined by performing locality tests~\cite{deringer2017machine,herbold2022hessian}. We found that typically 6~\AA{} to 10~\AA{} are sufficient for the majority of systems to construct a second generation MLP with state-of-the-art accuracy. The remaining interactions beyond this  cutoff radius are essential the long-range parts of dispersion and electrostatic interactions.
\par
In the second step we now further extend our method by focusing on the short-range atomic energies. As shown in Fig.~\ref{fig:example_4G-PINNP}, limitations in the description of the atomic environments can lead to contradictory training data reducing the overall accuracy of the potential. We therefore extend the input layer of the 4G-HDNNP short-range atomic NNs of atom $i$ by input neurons representing the electrostatic potentials $V_i^{j}$ arising from the charges of the neighboring atoms inside the local atomic environments. In order to avoid the cancellation of positive and negative contributions, we do not add the electrostatic potentials of all neighboring atoms. Instead, we make use of element-wise electrostatic potentials resulting in $j$ additional input neurons for systems containing $j$ elements. We note that both, the charges and the element-specific electrostatic potentials are descriptors allowing for a direct physical interpretation. Still, when used in the short-range energy atomic NNs, there is no constrained functional form like Coulomb's law, and as a consequence more complex and general relations between the input neurons and the atomic energies can be represented. 
\par
The structure of the resulting electrostatically embedded ee4G-HDNNP is shown schematically in Fig.~\ref{fig:extented_4G} for a binary example system. 
The total energy consists of an electrostatic part $E_{\mathrm{elec}}$ with infinite range, a short-range contribution $E_{\mathrm{short}}$, which also includes non-local electronic structure information, and the two-body contribution $E_{\mathrm{2b}}$,
\begin{equation}
    E_{\mathrm{total}}(\mathbf{R},\mathbf{Q},\mathbf{V}) = E_{\mathrm{elec}}(\mathbf{R},\mathbf{Q})+E_{\mathrm{short}}(\mathbf{R},\mathbf{Q},\mathbf{V})+E_{\mathrm{2b}}(\mathbf{R}),
\end{equation}
where $\mathbf{R}$ represents the atomic coordinates, $\mathbf{Q}$ the atomic partial charges, and $\mathbf{V}$ the electrostatic potentials. In the following paragraphs, we will now discuss the construction and the training of the electrostatic and the short-range energy contributions, while the two-body energy is calculated according to Eq.~\ref{eq:2body}.
\par
For the determination of the charges and the resulting electrostatic energies and forces we use the approach of the 4G-HDNNP framework. Specifically, the electrostatic energy $E_{\mathrm{elec}}$ is computed using a set of Gaussian broadened charges according to
\begin{equation}
       E_{\mathrm{elec}} = \sum_{i=1}^{N_{\mathrm{atoms}}}\sum_{j<i}^{N_{\mathrm{atoms}}} \frac{\erf{\frac{R_{ij}}{\sqrt{2} \gamma_{ij}}}}{R_{ij}} Q_i Q_j + \sum_{i=1}^{N_{\mathrm{atoms}}} \frac{Q_{i}^{2}}{2 \sigma_i \sqrt{\pi}}
\end{equation}
with 
\begin{equation}
    \gamma_{ij} = \sqrt{\sigma_i^2 + \sigma_j^2} \quad. \label{eq:gausscharge}
\end{equation}
and the $\sigma_{i}$ representing the widths of the Gaussian charge densities. The electrostatic energy for the periodic case can be calculated by Ewald summation~\cite{ewald1921ewald}.
Employing a charge equilibration scheme~\cite{P1448}, the partial charges are determined based on the minimization of 
\begin{equation}
    E_{\mathrm{Qeq}}= \sum_{i=1}^{N_{\mathrm{atoms}}} (\chi_{i}Q_{i}+\frac{1}{2}\eta_{i}Q_{i}^{2})+E_{\mathrm{elec}} \label{eq:eqeq}
\end{equation}
with respect to the charges, where $\chi_{i}$ and $\eta_{i}$ denote the atomic electronegativities and hardnesses. Like in 4G-HDNNPs, the electronegativites are constructed as local environment-dependent quantities, which are expressed by atomic NNs. These electronegativities and element-dependent hardnesses are optimized simultaneously during the training to reproduce the reference charges obtained from electronic structure calculations. To ensure the invariance of the atomic electronegativities under translation, rotation and permutation, a set of ACSFs is used to describe the local atomic environments, which depend on all neighboring atomic positions within a cutoff sphere of radius $R_{\mathrm{c}}$. All electronegativity atomic NNs have the same architecture and weights for a given element.
\par
The minimization of $E_{\mathrm{Qeq}}$ in Eq.~\ref{eq:eqeq} yields a set of linear equations
\begin{equation}
    \frac{\partial E_{\mathrm{Qeq}}}{\partial Q_{i}} = 0, \forall i =1,..,N_{\mathrm{atoms}} \implies \sum_{j=1}^{N_{\mathrm{atoms}}} A_{ij}Q_{j}+\chi_{i} = 0 
\end{equation}
in which the elements of the matrix $\textbf{A}$ are given by
\begin{equation}
    A_{ij} = 
        \begin{cases}
            J_i + \frac{1}{\sigma_i \sqrt{\pi}}, & \text{if}\ i=j \\
            \frac{\erf{\frac{R_{ij}}{\sqrt{2} \gamma_{ij}}}}{R_{ij}}, & \text{otherwise}
        \end{cases}
\end{equation}
These equations are solved under the constraint of conserving the total charge $Q_{\mathrm{tot}}$ of the system yielding
\begin{equation}\label{eq:eem}
    \begin{pmatrix}
        \mbox{\Large{$\M{A}$}}
        & \rvline & 
        \begin{matrix}
            1 \\ \vdots \\ 1
        \end{matrix}\\
        \hline
        \begin{matrix}
            1 & \hdots & 1
        \end{matrix}
        & \rvline &
        \begin{matrix}
            0
        \end{matrix}
    \end{pmatrix}
    \begin{pmatrix}
        Q_1 \\ \vdots \\ Q_{N_{\mathrm{atoms}}} \\ \hline \lambda
    \end{pmatrix}
    =
    \begin{pmatrix}
        -\chi_1 \\ \vdots \\ -\chi_{N_{\mathrm{atoms}}} \\ \hline Q_{\mathrm{tot}}
    \end{pmatrix} \quad .
\end{equation}
The electronegativity NN weights and the hardness values are both adjusted during the training process such that the error between reference and predicted atomic charges is minimized. In this study, the Hirshfeld charge partitioning scheme~\cite{P0416} is chosen but any other method can also be used in principle~\cite{verstraelen2016minimal,henkelman2006fast,heidar2017information}. Once the electronegativity atomic NNs have been trained, the electrostatic energies and forces can be computed as detailed in Ref.~\cite{P5932}.
\par 
In contrast to the short-range atomic energies in 4G-HDNNPs, which only depend on the charge on the central atom and a set of ACSFs as geometric descriptors of the environment, the most important extension of ee4G-HDNNPs is the refined description of the local electronic structure and its implications for local bonding by introducing a set of descriptors including non-local information. This information is represented by the element-resolved electrostatic potentials $\{V_{i}^{j}\}$ experienced by atom $i$ due to the Gaussian charges of all atoms of element $j$ within the cutoff sphere, which in turn depend on the entire system. Using separate input neurons for the electrostatic potential arising from all atoms of a given element ensures that no cancellation of positive and negative contributions of neighboring atoms can occur in case of predominantly ionic systems. 
\par
The expression of the descriptor $V_{i}^{j}$ in ee4G-HDNNPs has the form
\begin{equation}
     V_{i}^{j} = \sum_{k=1}^{N_{\mathrm{neig},j}}\frac{\mathrm{erf}(\frac{R_{ik}}{\sqrt{2}\gamma_{ik}})Q_{k}}{R_{ik}}f_{\mathrm{cut}}(R_{ik}) \quad ,
\end{equation}
i.e., $V_{i}^{j}$ can be considered as the total short-ranged electrostatic potential acting on the central atom $i$ arising from all $N_{\mathrm{neig},j}$ neighboring atoms, which belong to element $j$. The cutoff function is included to make sure that the electrostatic potential and its derivative decays to zero at the cutoff radius $R_{\mathrm{c}}$, which we choose here to be the same as the cutoff for the atomic energies and electronegativities. It provides a fingerprint of the element-specific local charge density around the reference atom facilitating the description of changes in the atomic interactions due to changes in chemical bonding, including those resulting from distant changes in the system, e.g. due to long-range charge transfer.
\par
In the training process of the ee4G-HDNNP, first the electronegativity atomic NNs and atomic hardnesses are trained. In a second step, the short-range energy function $E_{\mathrm{short}}$, which now includes detailed electronic information, is trained to predict the remaining part of the total energy and the atomic forces after removing the electrostatic energy $E_{\mathrm{elec}}$, the electrostatic forces and the respective two-body interactions from the reference electronic structure energy and forces in order to avoid the double counting of energy and force contributions. For instance, the target short-range energy for training the ee4G-HDNNP can be written as 
\begin{eqnarray}
    E_{\mathrm{short}}^{\mathrm{target}}({\textrm{ee4G-HDNNP}})&=&E_{\mathrm{ref}}-E_{\mathrm{elec}}-E_{\mathrm{2b}}\nonumber \\
    &=&\sum_{i=1}^{N_\mathrm{atoms}}E_{i}(\mathrm{\mathbf{G}}_{i},Q_{i},\mathrm{\mathbf{V}}_{i})
\end{eqnarray}
The $\mathrm{\mathbf{G}}_{i}$ and $\mathrm{\mathbf{V}}_{i}$ represent the vectors of symmetry functions and element-dependent electrostatic potentials acting on atom $i$, respectively. The total number of input neurons of the short-range atomic NNs is equal to the sum of the number of ACSFs, the number of elements regarding the electrostatic potentials plus one for the atomic charge of the central atom itself (see Fig.~\ref{fig:extented_4G}). Introducing the electrostatic potential descriptors into the model does not cause significant extra computational costs due to the fact that the electrostatic potential and its gradient with respect to the atomic positions can be directly computed from the physical quantities already calculated in the framework of the original 4G-HDNNP such as charges and their derivatives with respect to the atomic positions.
\par
To systematically investigate the relevance of each modification for the accuracy of the PESs, next to the ee4G-HDNNP also a conventional 4G-HDNNP and a 4G-HDNNP$+E_{\mathrm{2b}}$ that only includes additional two-body terms are developed. In this case the target short-range energy for the training step is given by
\begin{eqnarray}
     E_{\mathrm{short}}^{\mathrm{target}}({\textrm{4G-HDNNP}+\mathit{E}_{2b}})  &=&E_{\mathrm{ref}}-E_{\mathrm{2b}}-E_{\mathrm{elec}}\nonumber \\
     &=&\sum_{i=1}^{N_{\mathrm{atoms}}} E_{i}(\mathbf{G_i},Q_i)  \quad.
\end{eqnarray}
Hence, the only difference between the ee4G-HDNNP and the 4G-HDNNP+$E_{\mathrm{2b}}$ is the short-range energy.

\section{Computational Details}

\subsection{DFT calculations}

All DFT calculations have been carried out using the electronic structure code FHI-aims~\cite{blum2009ab} (version: 200112\_2) with ``light'' settings. The PBE functional has been employed to describe electronic exchange and correlation~\cite{perdew1996generalized}. The SCF convergence criteria of the total energy, the volume integrated root mean square of the charge density and the sum of eigenvalues have been set to $10^{-5}$ eV, $10^{-4}$ $e$, and $10^{-2}$ eV, respectively.  
\par
The dataset consists of Na$_{n}$Cl$_{n}$ and Na$_{n}$Cl$_{n+1}^{-}$ clusters with $n=16$ and $n=24$ and the data generation can be divided into two stages. In the first stage, 10 smaller and 5 larger clusters for each charge state have been chosen to perform Born—Oppenheimer molecular dynamic simulations at 1000~K for 5000 steps with a time step of 1.0 fs. The configurations of every 5th step in the trajectories have been included in the dataset. The initial structures for these trajectories have been obtained from minima hopping using BigDFT~\cite{genovese2011daubechies} and then re-optimized with FHI-aims using the force threshold of 0.01 eV/\AA. A Nos\'{e}-Hoover thermostats was applied to run simulations in the canonical ensemble, and the effective mass was set to 1700 cm$^{-1}$. Apart form that, the paths of the geometry optimizations have also been included in the dataset in order to have a sufficient sampling in the lower energy region of configuration space. 
\par
In the second stage extrapolating structures that are not well sampled in the configuration space at low temperature have been identified by running short MD simulations driven by a preliminary ee4G-HDNNP. Moreover, local minima, which have been generated by the Coulomb Lennard Jones empirical force field~\cite{faro2010lennard} using Artificial Bee Colony algorithms~\cite{zhang2015abcluster} and relaxed with the preliminary potential have also been included in the dataset if the optimized geometry exhibits large structural deviations compared to the DFT result. 
\par
The final dataset consists of 33,592 structures including 10,627 Na$_{16}$Cl$_{16}$ clusters, 5,822 Na$_{24}$Cl$_{24}$ clusters, 11,217 Na$_{16}$Cl$_{17}^-$ clusters, and 5,927 Na$_{24}$Cl$_{25}^-$ clusters. 31,253 of these clusters have been obtained in the first stage, and 2,339 in the second stage.
\par 
For the transferability tests to periodic NaCl systems, an 8 $\times$ 8 $\times$ 8 k-point grid has been used to compute the equation of state for the NaCl-FCC structure, while a 4 $\times$ 4 $\times$ 4 k-point grid has been chosen for performing Born—Oppenheimer molecular dynamic simulations of the NaCl melt.  
\subsection{Construction of the neural network potentials}

The HDNNPs reported in this work have been constructed using the program RuNNer~\cite{behler2017runner,behler2015constructing,behler2017first}.
 Atom-centered symmetry functions~\cite{behler2011atom} have been used for the description of the atomic environments within a spatial cutoff radius of 12 Bohr $\approx$ 6.35 \si{\angstrom}. The same parameters of the symmetry functions and the same atomic NN architectures have been used for all HDNNPs. The functional forms of the symmetry functions are given in Ref.~\cite{behler2011atom}, while the specific parameters we used are given in the supplementary information. For computing the electrostatic energies, the widths of the Gaussian charges have been set to 1.54 and 0.99 \si{\angstrom} for Na and Cl, respectively. The atomic NNs consist of an input layer with the same number of ACSFs for Na and Cl, two hidden layers with 15 neurons each, and an output layer with one neuron providing either the atomic short range energy or the electronegativity. Forces have been obtained as analytic energy derivatives. As activation functions in the hidden layers and the output layer the hyperbolic tangent and the linear function have been selected, respectively. A global, adaptive, extended Kalman filter~\cite{blank1994adaptive} was used to optimize the neural network weights. In addition, the L2 regularization of weights during the training of the electrostatic part and short-range part was set to $10^{-4}$ and $10^{-6}$, respectively. However, no regularization was applied to the electrostatic part of the 3G-HDNNP in order to prevent large gradients for the weight update during the training. 
\par
90\% of the reference data were used for training the potentials while the remaining 10\% of the data points were used as an independent test set to confirm the reliability of PESs and to detect possible over-fitting. Energies and forces were used for training the short-range atomic NNs. 

\subsection{Minima Hopping algorithm}

The parameters for minima hopping follow the notation in Ref.~\cite{goedecker2004minima} and were set to $\alpha_{1}=0.9,\alpha_{2}=1.1,\beta_{1}=1.1,\beta_{2}=1.1,\beta_{3}=0.909$, and $E_{\mathrm{diff}}=2\times10^{-3}$ Ha. 
Some extensions to the original minima hopping were implemented to increase the stability of the simulations.
A safe threshold for the temperature was introduced to limit the maximum temperature of the escape step of the minima hopping method. 
Additionally, a fixed number of time steps was used for the trajectory which was adjusted dynamically. 
The safe threshold for the temperature was set to $1000$\,K. 
Once a temperature above this threshold was reached, the number of time steps was increased by a factor of 2 until a value larger than a predefined maximum of $n_\text{max}=6400$ was reached.
After this, the temperature was allowed to increase beyond the safe threshold.
The temperature and the number of time steps was decreased again by a factor of 2 in case they were above the safe threshold or the initial number of time steps when a new minimum was accepted.
With these modifications very high temperatures can be effectively avoided.
In addition, the minimum temperature and initial number of time steps were set to $500$\,K and $100$ during the simulations to increase the efficiency.
The initial temperature of the MD simulations was set to $100$\,K, and a time step of $0.5$\,fs has been used. 
The force threshold for the geometry relaxations has been set to $0.01$\,eV/\si{\angstrom}. 
Softening with 40 iterations was used to adjust the initial velocities towards low-curvature directions of the PES. 
The parameters $d$ and $\alpha$ were set to $0.1$ and $0.015$, respectively. 
More detailed information about velocity softening can be found in Ref.~\cite{Sicher2011,schonborn2009performance}. 
The RMSD of the predicted minima with respect to DFT optimized structures was calculated using the tool developed in Ref.~\cite{finkler2020funnel}. 
More detailed information about minima hopping can be found in Ref.~\cite{goedecker2004minima}.

\section{Results and Discussion}

\subsection{Quality of the HDNNPs}

\begin{table}[]
    \footnotesize
    \centering
        \caption{\textbf{The training and test set errors of different HDNNPs.} Root mean squared errors (RMSE) of the NaCl binding energies (in meV/atom), forces (in meV/\si{\angstrom}) and charges (in m$e$) of the 3G-HDNNP, 4G-HDNNP, 4G-HDNNP+$E_{\mathrm{2b}}$ and ee4G-HDNNP for the structures in the training set. The respective errors of test set are given in parentheses. Note that the RMSE of the charges is the same for all fourth-generation HDNNPs due to the identical charge training procedure.}
    \label{tab:RMSE}
    \begin{tabular}{c c c c}
         Method & Energy & Forces  & Charges \\
        \hline\hline
         3G-HDNNP & 2.156 (2.205)  & 124.52 (124.09) & 5.854 (5.870) \\
         4G-HDNNP &  1.779 (1.796) & 112.20 (110.33)& 6.904 (6.919)\\
         4G-HDNNP+$E_{\mathrm{2b}}$ & 1.391 (1.381) & 68.33 (68.21)& 6.904 (6.919)\\
         ee4G-HDNNP  & 1.250 (1.267)  & 60.20 (60.53) & 6.904 (6.919)\\
    \end{tabular}
\end{table}

The DFT reference dataset consists of 33,592 neutral and negatively charged sodium chloride clusters of composition Na$_{n}$Cl$_{n}$ and Na$_{n}$Cl$_{n+1}^{-}$ with $n$=16 and 24 that have been generated in molecular dynamics (MD) simulations and geometry relaxations. The resulting structures cover an energy range of about 0.692 eV/atom.
\par
The first step to assess the accuracy of the potentials is to determine the root mean square error (RMSE) of the binding energies, atomic forces and atomic charges obtained with the 3G-HDNNP, 4G-HDNNP, 4G-HDNNP+$E_{\mathrm{2b}}$ and ee4G-HDNNP. These errors with respect to the DFT reference are compiled in Table~\ref{tab:RMSE} for the training and the test set.
\par
For the atomic charges the error of about 6.9$\cdot 10^{-3} e$ is the same for all fourth-generation methods, which share the same charge training procedure. As expected for this system, the 3G-HDNNP is performing even slightly better since overall in NaCl the charges predominantly depend on the local environment and are similar for all atoms of a given element. In this special case, including both neutral and negative clusters in the training set does not pose a severe problem for a local method like the 3G-HDNNP, since the total charge is closely correlated to the number of chlorine and sodium atoms. Nevertheless, the 3G-HDNNP shows the largest errors for the energies and forces, which can be ascribed to the short-range atomic energies depending only on the local geometric atomic environments described by the ACSFs. Subtle changes in the charge density, e.g. polarization, due to structural features outside the atomic environments cannot be accounted for in this method. This is different in the 4G-HDNNPs, in which the charges depend on the global structure and net charge of the system. Consequently, the 4G-HDNNP method provides smaller errors of the energies and forces than the 3G-HDNNP, in spite of slightly larger atomic charge errors. This is a result of the additional information included in form of the atomic partial charge in the short-range atomic NNs. Variations in this charge as a function of the structure of the system thus seem to assist the construction of more accurate short-range energies. Including empirical 2-body terms further reduces the errors, and in particular the forces show a strongly reduced error since, e.g., repulsive interactions are already included in the simple 2-body energy reducing the complexity of the remaining PES to be represented by the short-range energies. Finally, including additional descriptors representing the electrostatic potential of the environment in the ee4G-HDNNP yields overall the smallest errors of all investigated MLPs. The correlation plots of the energies, forces and charges for all potentials are provided in the supplementary information.

\subsection{Atomic forces on benchmark clusters}

\begin{figure}
    \centering
    \includegraphics[width =0.75\textwidth]{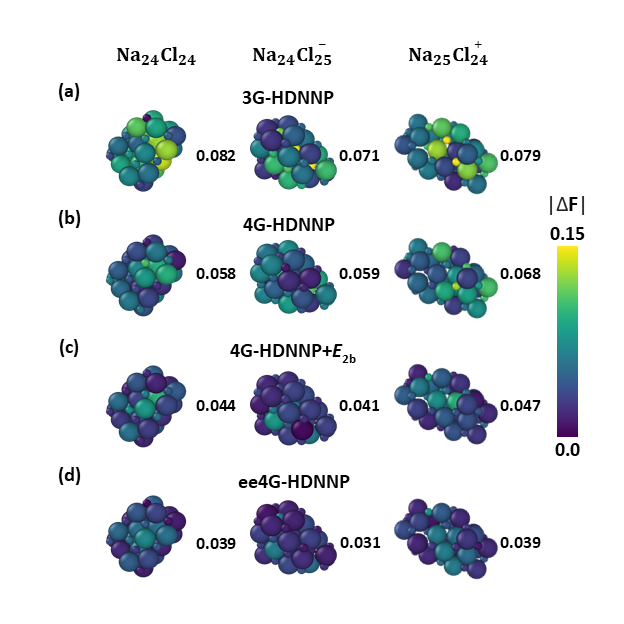}
    \caption{\textbf{Atomic force errors for neutral and charged NaCl clusters.} The colors represent the absolute errors of the atomic forces with respect to DFT in eV/\AA{} for Na$_{24}$Cl$_{24}$, Na$_{24}$Cl$_{25}^{-}$ and Na$_{25}$Cl$_{24}^{+}$ clusters as obtained with the 3G-HDNNP (a), 4G-HDNNP (b), 4G-HDNNP+$E_{\mathrm{2b}}$ (c) and ee4G-HDNNP (d). The numbers next to the clusters give the root mean squared errors of the atomic forces compared to DFT. The sodium and chlorine atoms are shown as large and small spheres, respectively.}
    \label{fig:4G_proof_of_concept}
\end{figure}

While the RMSE values of the binding energies, forces and charges provide an overview about the average accuracy of the potential, they do not yield detailed information about the quality of the PES for specific geometries, nor do they allow to predict the transferability of the HDNNPs beyond the training set, e.g. for clusters with different total charges. In order to investigate the accuracy of the individual atomic forces, the left column of Fig.~\ref{fig:4G_proof_of_concept} shows the errors of the atomic forces in a cluster of composition Na$_{24}$Cl$_{24}$ in the DFT-optimized geometry as obtained with the different HDNNPs. It can be clearly seen that the deviations of the atomic forces decay in the same sequence like the RMSEs from 3G-HDNNP via 4G-HDNNP and 4G-HDNNP+$E_{\mathrm{2b}}$ to ee4G-HDNNP. This decay can also be quantified by the RMSEs of the forces of all atoms in this cluster, which decrease from 0.082 eV/\AA{} to 0.039 eV/\AA{}. The same trend is also observed for the negatively charged Na$_{24}$Cl$_{25}^{-}$ cluster in the middle column, and also in this case the RMSE is reduced to less than half. Finally, to assess the transferability of the HDNNPs to charge states not included in the reference data, we have investigated the Na$_{25}$Cl$_{24}^{+}$ cluster shown in the right column of Fig.~\ref{fig:4G_proof_of_concept}. The force errors with respect to DFT are very similar to the other two clusters indicating a very good transferability to a positively charged cluster containing an excess sodium ion. 

In addition to the overall RMSE of the forces, which could arise either from similar errors of all forces or a few prominent outliers with particularly large errors, the absolute force errors of the individual atoms are also of high interest, in particular close to stationary points of the PES like local and global minima. The largest atomic force error of the clusters in Fig.~\ref{fig:4G_proof_of_concept} obtained with the 3G-HDNNP is about 0.17 eV/\si{\angstrom}. 
The 4G-HDNNP slightly reduces the largest force error to around 0.15 eV/\si{\angstrom}. The error is further reduced to 0.13 eV/\si{\angstrom} by including empirical two-body terms in the 4G-HDNNP+$E_{\mathrm{2b}}$. Finally the ee4G-HDNNP significantly decreases the maximum error to only about 0.08 eV/\si{\angstrom}, which is half of the error of the 3G-HDNNP although the overall RMSE of the training set of this potential is lower. This shows that the ee4G-HDNNP does not only yield a considerably improved description of the atomic interactions but also a more balanced force error distribution for all atoms.

\subsection{Structural analysis and energetic ordering of local minima}

Many applications require reliable energy and force predictions for a wide range of structures and bonding situations. To further explore the accuracy of the ee4G-HDNNP for different cluster geometries, we performed structure prediction using minima hopping~\cite{goedecker2004minima}. This is a particularly challenging test for potentials due to the typically very small energy differences between different local minima. Moreover, the MD simulations at elevated temperatures included in the minima hopping require a stable potential also in non-equilibrium situations.

\begin{figure*}
    \centering
    \includegraphics[width=1.0\textwidth]{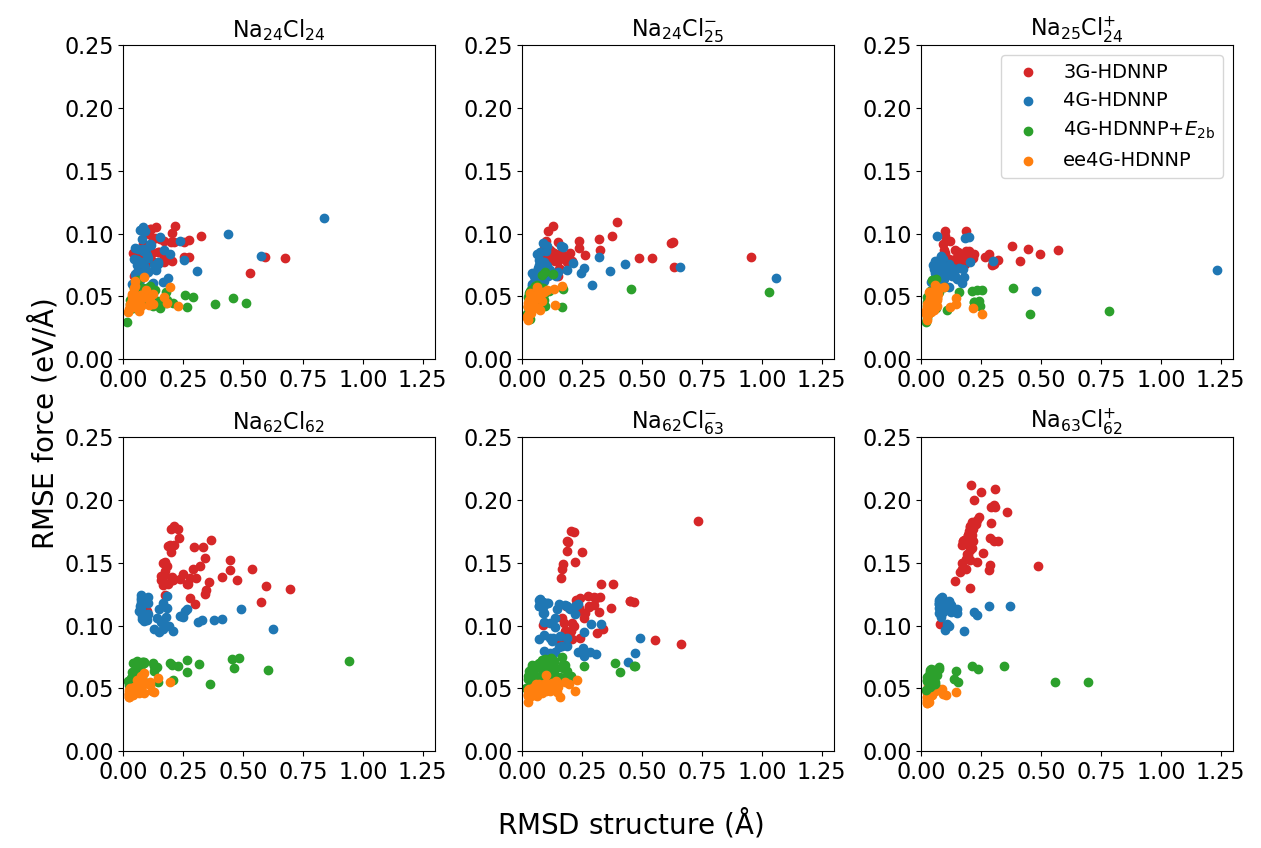}
    \caption{\textbf{Structural and force deviations of local minima obtained with different HDNNPs in comparison to DFT.} Shown are the correlation plots of the root mean squared errors (RMSE) of the forces and the root mean squared displacements (RMSD) of the atomic positions for a series of local minima predicted by the 3G-HDNNP, 4G-HDNNP, 4G-HDNNP+$E_{\mathrm{2b}}$ and ee4G-HDNNP with respect to DFT-optimized geometries for neutral, negatively and positively charged NaCl clusters of different size.}
    \label{fig:correlation_between_energy_and_RMSD}
\end{figure*}

For each run, we investigated the first 50 minima that have been obtained, which are not necessarily the 50 lowest minima to make sure that different structural patterns are present. Then, to have a fair comparison between the 3G-HDNNP, 4G-HDNNP, 4G-HDNNP+$E_{\mathrm{2b}}$ and DFT, we re-optimized these local minima predicted by the ee4G-HDNNP with the other methods, i.e., DFT, the 3G-HDNNP, the 4G-HDNNP and the 4G-HDNNP+$E_{\mathrm{2b}}$ to determine the closest local minima of the respective PESs.  
We have explored 50 minima each for neutral and charged sodium chloride clusters of composition Na$_{n}$Cl$_{n}$, Na$_{n}$Cl$_{n+1}^{-}$ and Na$_{n+1}$Cl$_{n}^{+}$ with $n=24$ and $n=62$ resulting in total in 300 geometries. The results of all clusters are given in Fig.~\ref{fig:correlation_between_energy_and_RMSD}. Again we note that the positively charged clusters, which have not been included in the training set, allow us to test the transferability of the PESs to different charge states.
\par
First, the different HDNNPs have been investigated by determining the structural deviations, i.e., the root mean squared displacement (RMSD), of the predicted minima with respect to the DFT optimized structures. 
In addition, we also computed the DFT forces for the minimum geometries predicted by the HDNNPs to determine the force RMSE of each minimum structure with respect to DFT.
Fig.~\ref{fig:correlation_between_energy_and_RMSD} shows the resulting correlation plots of structure and force errors for all cluster sizes and charges for the different HDNNPs. A clear trend can be observed in that in all cases the errors of both, positions and forces, with respect to DFT decrease in the order of the 3G-HDNNP, 4G-HDNNP, 4G-HDNNP+$E_{\mathrm{2b}}$ and ee4G-HDNNP with the latter providing the best agreement with DFT in all cases. 
While the RMSDs do not show a significant difference between the smaller and the larger clusters, in general the forces are more accurate for the smaller clusters, which are easier to describe even by a local method like the 3G-HDNNP relying on environment-dependent charges. Hence, the performance of the 3G-HDNNP and the 4G-HDNNP, which both include short-range and electrostatic energies, is similar for the smaller clusters, while the 4G-HDNNP as a global method shows a better accuracy for the larger clusters.
\par
We note that for the 3G-HDNNP-based geometry optimizations of two of the clusters of composition Na$_{62}$Cl$_{62}$ converged to unphysical geometries with too short bond lengths due to extrapolation, as no clusters of this size have been included in the training set. We did not observe such problems for any other method, and in particular the inclusion of 2-body terms efficiently prevents too close atomic encounters.
For all clusters, the inclusion of 2-body interactions generally reduces the force errors to about the order of magnitude found for the smaller clusters, which is a clear evidence for the higher transferability of the potential, but without strongly improving the structures in most cases. Finally, the ee4G-HDNNP performs best of all methods, for all cluster sizes and charges, with clearly decreased structural and force deviations of at most 0.26~\si{\angstrom} and 0.07~eV/\si{\angstrom}, respectively, which can be ascribed to the additional information about the electronic structure. Noteworthy, there is no difference in the accuracy of the ee4G-HDNNP for structures and forces between the small and large clusters, underlining the excellent transferability, which is also confirmed by the accurate predictions for the positively charged clusters not included in the training set at all. We found the ee4G-HDNNP to provide stable geometry optimizations and trajectories in spite of formal extrapolations in particular for the larger clusters. More than 95$\%$ of all minima have a force RMSE below 0.06 eV/\si{\angstrom}, which is also about the training set RMSE of the forces of the ee4G-HDNNP, and 85$\%$ of minima show an RMSD of less than 0.1 \si{\angstrom}.
\begin{figure*}
    \centering
    \includegraphics[width=1.0\textwidth]{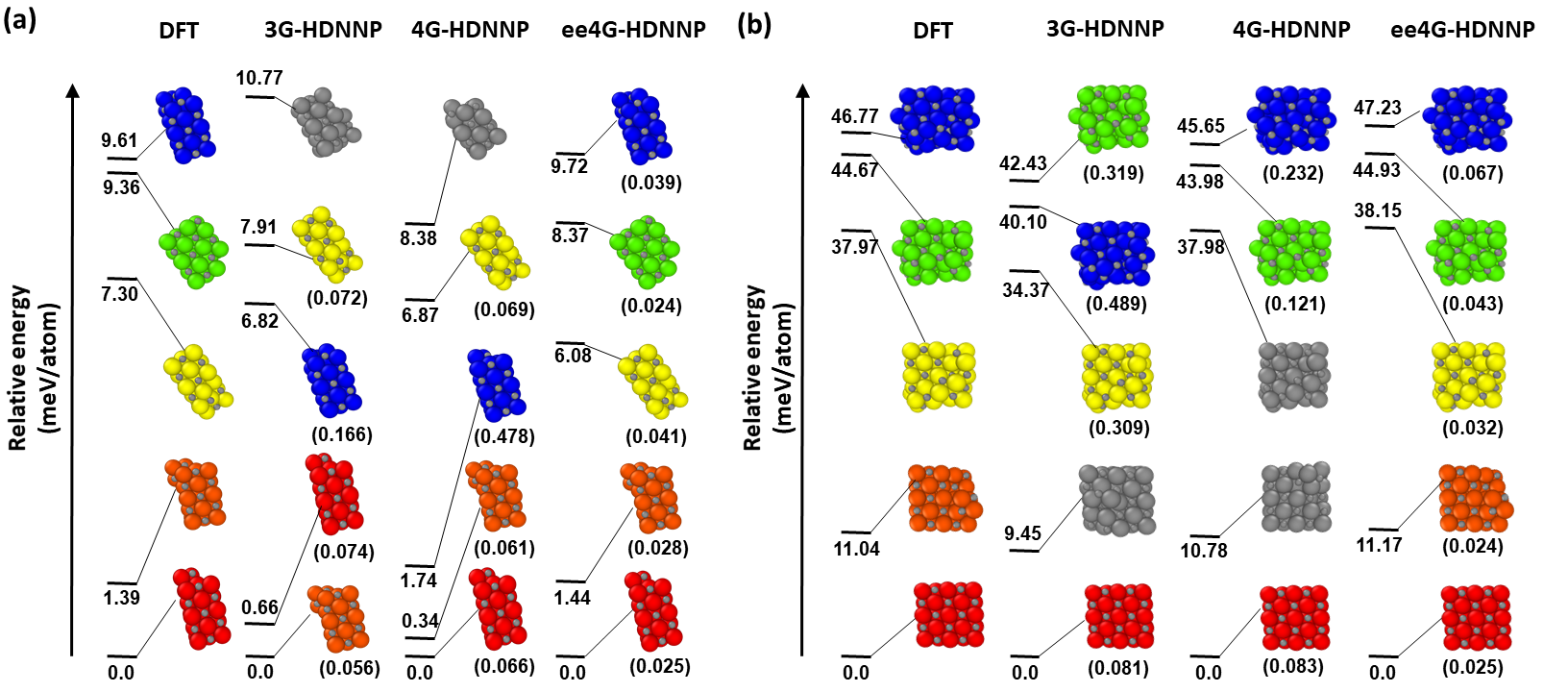}
    \caption{\textbf{Energetic ordering of local minima of positively charged clusters predicted by different methods.} Shown are the five lowest energy minima of Na$_{25}$Cl$_{24}^{+}$ (a) and five representative structures of Na$_{63}$Cl$_{62}^{+}$ covering the energy range from the lowest to the highest energy minimum predicted by DFT (b). The numbers give the relative energy in meV/atom with respect to the lowest energy structure predicted by the 3G-HDNNP, 4G-HDNNP and ee4G-HDNNP. The numbers in brackets indicate the root means square displacement (RMSD) in \si{\angstrom} with respect to the DFT optimized geometry, which has been used as initial structure to determine the minima of the respective HDNNP type. The colors are used to highlight the relation to the original DFT geometries.}
    \label{fig:five_lowest_minima}
\end{figure*}
\par
Besides the investigation of the structural and force errors of the local minima, we have also studied the energetic ordering of these structures, which is a stringent test due to the often tiny energy differences. We focus our discussion here on the positively charged clusters, which are most interesting as they have not been included in the training set. Fig.~\ref{fig:five_lowest_minima}a shows the relative energies of the five lowest DFT minima among the Na$_{25}$Cl$_{24}^{+}$ clusters investigated in Fig.~\ref{fig:correlation_between_energy_and_RMSD}. These clusters have then been re-optimized by the 3G-HDNNP, the 4G-HDNNP and the ee4G-HDNNP. The resulting structural changes in form of the RMSD are shown in parentheses below the clusters. All structures have been colored to show their relation to the DFT ancestor geometries, with grey structures indicating geometries originating from DFT parent structures different from the five given on the left. In spite of the very small energy differences of at most a few meV/atom and often less, the ee4G-HDNNP shows an energetic ordering in excellent agreement with DFT, both qualitatively and quantitatively, while the order is in some cases different for the 3G-HDNNP and the 4G-HDNNP. Still, the 4G-HDNNP is able to identify the global minimum and the lowest local minimum correctly. To probe the higher-energy clusters, Fig.~\ref{fig:five_lowest_minima}b shows five minima of composition Na$_{63}$Cl$_{62}^{+}$ covering the full energy range of the 50 available minima of this cluster. In this case all HDNNPs types provide the same global minimum as DFT, but again only the ee4G-HDNNP provides the same energetic ordering of all structures with energy differences agreeing to within about 1 meV/atom with DFT. All these results demonstrate both, the excellent transferability and accuracy, of the ee4G-HDNNP to larger systems and even different charge states that are not included in the dataset.

\subsection{Transferability of HDNNPs to periodic systems}

%
\begin{figure}
    \centering
    \includegraphics[width=0.5\textwidth]{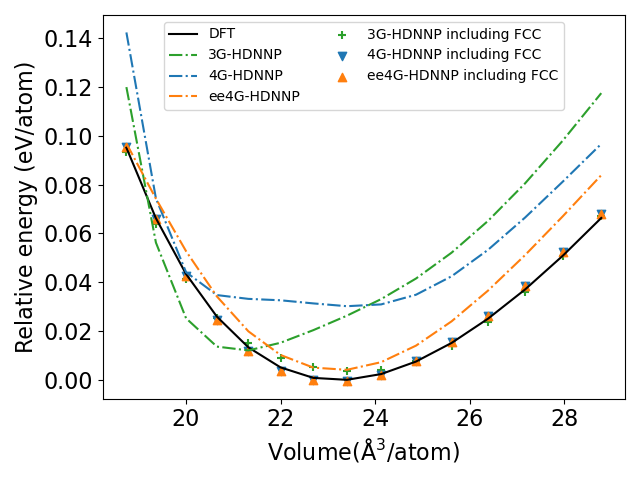}
    \caption{\textbf{Equation of state of the FCC NaCl crystal structure obtained by different potentials.} Shown is the relative energy per atom of the cubic NaCl structure as a function of the atomic volume. Three potentials, 3G-HDNNP, 4G-HDNNP and ee4GHDNNP, trained to cluster data only and trained to a dataset including periodic fcc NaCl structures have been investigated. The energies are given relative to the minimum energy of the FCC structure computed by DFT.}
    \label{fig:EOS}
\end{figure}
\begin{table}[]
    \centering
\caption{\textbf{Equilibrium lattice constant a$_{0}$ and bulk modulus B$_{0}$ of an FCC NaCl crystal.} The properties are computed from the Murnaghan equation of state based on DFT, the 3G-HDNNP, the 4G-HDNNP and the ee4G-HDNNP. }
    \begin{tabular}{c c c c c}
    property & DFT & 3G-HDNNP & 4G-HDNNP &  ee4G-HDNNP \\
    \hline\hline
 a$_{0}$(\si{\angstrom})  & 5.706 &5.555 & 5.602 & 5.708 \\
             B$_{0}$(GPa) & 23.32 &  43.17 & 28.73 &  25.79 \\
    \hline
    \end{tabular}
    \label{tab:Equilibrium bond}
\end{table}
Finally, we test the transferability of the potentials, which have been constructed using small cluster data only, to the periodic NaCl FCC crystal structure. In the periodic crystal, the atomic environments corresponding to a cutoff of 6.35~\AA{} contain approximately 32 atoms for the equilibrium lattice constant and we note that such bulk-like fragments are not present in the cluster training set. Fig.~\ref{fig:EOS} shows the equation of state computed by DFT and by the 3G-HDNNP, the 4G-HDNNP and the ee4G-HDNNP. The equilibrium lattice constants and bulk moduli computed by a fit to the Murnaghan equation of state are given in Table~\ref{tab:Equilibrium bond}. While all HDNNPs provide smooth energy curves with single minima, only the curve of the ee4G-HDNNP is acceptably close to the DFT reference, with an essentially identical equilibrium lattice constant and a cohesive energy differing only by about 4 meV/atom. 
The calculated RMSEs for the 3G-HDNNP, the 4G-HDNNP and the ee4G-HDNNP in the plotted lattice constant range are 30.7, 27.3, 9.5 meV/atom (mean absolute errors: 27.1, 25.0 and 8.5 meV/atom), respectively.
This is significantly larger than the RMSEs of the HDNNPs for the training clusters given in Table~\ref{tab:RMSE}, because in the energy versus volume curves interatomic distances vary strongly far beyond the close-to equilibrium cluster geometries included in the training set. Notably, the partial charges predicted by 4G-HDNNPs based on the charge equilibration scheme also match DFT results as displayed in the supplementary information. 
\par
In order to confirm that these results indeed can be ascribed to the different transferabilities of the potentials and are not caused by intrinsic limitations of the methods, we have augmented the cluster datasets by including periodic NaCl crystals with different lattice constants from 4.55 \si{\angstrom} to 6.83 \si{\angstrom}. As can be seen in Fig.~\ref{fig:EOS}, the revised potentials including the FCC crystal structures are all able to represent the DFT curve with very high accuracy.

To further examine the transferability of the ee4G-HDNNP to periodic systems, the structural properties of the NaCl melt have been investigated. It should be emphasized that such analyses are particularly interesting since the dataset does not contain any periodic structures and there is no unique correlation between atomic positions and potential energy due to inclusion of multiple charge states.       

We generated the molten salt by performing NVT simulations driven by ee4G-HDNNP for the cubic super cell at 1400~K, which contains 32 ion pairs with a fixed volume of 2.248 nm\textsuperscript{3} taken from Ref.~\citep{tovey2020dft}. The total MD simulation time includes an initial 0.5 ps equilibration, and the following 100 ps for performing radial distribution function (RDF) analysis. For comparison we generated also a DFT RDF by computing a 10 ps Born—Oppenheimer MD trajectory after an equilibration of the same length. The RDF analyses were performed using the R.I.N.G.S program~\citep{le2010ring}.
\begin{figure}
    \centering
    \includegraphics[width=0.5\textwidth]{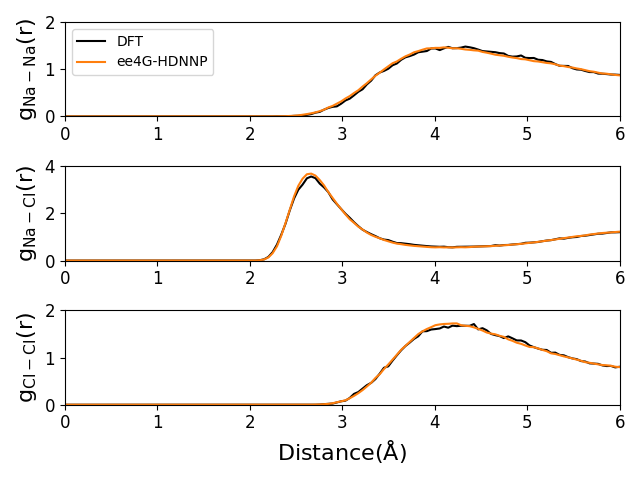}
    \caption{\textbf{Radial distribution functions of the NaCl melt.} Na-Na, Na-Cl and Cl-Cl RDFs computed using ee4G-HDNNP-driven MD simulations at 1400~K, compared with reference DFT Born—Oppenheimer MD data.}
    \label{fig:RDF_1400K}
\end{figure}
Fig.~\ref{fig:RDF_1400K} shows that the ee4G-HDNNP accurately reproduces the heights and the positions of the first peak of different RDFs in comparison to the DFT results, which are also consistent to the results from Ref~\citep{tovey2020dft}. This clearly illustrates that ee4G-HDNNP exhibits a convincing tranferability from non-periodic to periodic structures. 

\section{Conclusions \label{sec:Conclusions} }

In this work, we have developed an electrostatically embedded fourth-generation high-dimensional neural network potential (ee4G-HDNNP), which is based on the 4G-HDNNP with two extensions. The first is an improvement of the short-range atomic energies by the introduction of a set of additional descriptors, which can be interpreted as the element-resolved electrostatic potential arising from all neighboring atoms within a cutoff sphere to provide additional information about the electronic structure in the system. These electrostatic potentials make use of atomic charges obtained in the global charge equilibration scheme and are thus able to take non-local phenomena like long-range charge transfer and subtle changes in the electronic structure into account. 
In addition, the transferability of the potential has been enhanced by including empirical two-body potentials inspired by Tosi-Fumi model, which take Pauli repulsion and dispersion interactions into account in an approximate way. The required parameters can be obtained from the binding curves of free dimers.
\par 
The capabilities of the ee4G-HDNNP have been demonstrated for a dataset containing small neutral and negatively charged NaCl clusters with large structural diversity, including a systematic comparison with the 3G-HDNNP, 4G-HDNNP and 4G-HDNNP with only two body potentials (4G-HDNNP+$E_{\mathrm{2b}}$). All our analyses reveal that the accuracy and transferability of the HDNNPs is gradually improved from the 3G-HDNNP via the 4G-HDNNP and the 4G-HDNNP+$E_{\mathrm{2b}}$ to the ee4G-HDNNP. The ee4G-HDNNP is able to predict the properties of larger clusters and of positively charged clusters not included in the training set with DFT quality such as the structure of local minima and their subtle energetic ordering. Finally, the transferability of the ee4G-HDNNP in case of extrapolation beyond the training set has been shown in the periodic structures including the NaCl crystal and the melt based on the dataset containing only cluster structures. Compared to 4G-HDNNPs there is only a minor increase in the computational costs since the atomic charges and their derivatives required for the electrostatic potential descriptors are readily available. Physically interpretable descriptors like the atomic charges and electrostatic potentials improve the description of the atomic environments and are thus suitable to overcome current limitations in the resolution of the atomic energies due to the incompleteness of local atomic environments represented by two and three-body structural descriptors~\cite{pozdnyakov2020incompleteness}.
\par
In terms of scalability of the model, the charge equilibration step in the framework of the 4G- and ee4G-HDNNPs incurs additional computational cost compared to second generation methods. Despite the availability of highly optimized linear solvers~\cite{demmel1989lapack,choi1996scalapack}, the 4G- and ee4G-HDNNPs could still become demanding for simulating very large systems beyond tens of thousands of atoms.
From our investigation on the performance of 4G- and ee4G-HDNNPs as shown in supplementary information, the bottleneck of the performance is mainly attributed to the calculation of electrostatic energies and forces of a periodic system for systems containing up to a few thousand atoms. 
The performance can hence be significantly improved by using alternatives, such as particle-particle-particle mesh~\cite{greengard1988rapid} and fast multipole methods~\cite{greengard1987fast}, which can achieve quasi-linear scaling.
We expect that these methods combined with advanced iterative approaches for solving the linear equation system~\cite{leven2019inertial, o2020fast} could significantly accelerate the computation of the partial charges and their derivatives has been demonstrated in the framework of reactive force fields~\cite{leven2021recent}.

\section{Acknowledgements}
We are grateful for financial support by the Deutsche Forschungsgemeinschaft (DFG) (BE3264/13-1, project number 411538199) and the Swiss National Science Foundation (SNF) (project number 182877 and NCCR MARVEL).
The calculations presented here were performed in Göttingen (DFG INST186/1294-1 FUGG, project number 405832858). The NaCl minimum geometries provided for our DFT calculations of the initial training set have been computed at sciCORE (http://scicore.unibas.ch/) scientific computing center at University of Basel and the Swiss National Supercomputer (CSCS) under project s963D/C03N05.\\

\section{Data availability}

The datasets and analyses presented in this work are available from the corresponding author on request. In addition, the data will be made available in Materials Cloud once the manuscript has been accepted for publication.
\section{Code availability}

All DFT calculations were performed using FHI-aims (version 200112\_2)~\cite{blum2009ab}. The neural network potentials have been constructed using the program RuNNer~\cite{behler2017runner,behler2015constructing,behler2017first}, which is available free of charge under the GPL3 license at \url{https://www.uni-goettingen.de/de/software/616512.html}.

\bibliography{literature}

\end{document}